\documentclass[12pt,preprint]{emulateapj}

\def\kms{km~s$^{-1}$}

\slugcomment{Updated Oct 22/08.  Matches version accepted by ApJ.}

\shorttitle{Late-time Optical Emission from SN 1979C}
\shortauthors{Milisavljevic et al.}

\begin{document}

\title{The Evolution of Late-time Optical Emission from SN 1979C \\
~}

\author{Dan Milisavljevic, Robert A.\ Fesen}
\affil{6127 Wilder Lab, Department of Physics \& Astronomy, Dartmouth
  College, Hanover, NH 03755 }
\and
\author{Robert P.\ Kirshner, Peter Challis}
\affil{Harvard-Smithsonian Center for Astrophysics, 60 Garden Street,
  Cambridge, MA 02138}

\begin{abstract}

Optical spectra of the bright Type II-L supernova SN 1979C obtained in April
2008 with the 6.5~m MMT telescope are compared with archival late-time spectra
to follow the evolution of its optical emission over the age range of 11 to 29
years.  We estimate an H$\alpha$ flux decrease of around 35\% from 1993 to 2008
but noticeable increases in the strength of blueshifted emission of forbidden
oxygen lines. While the maximum expansion of the broad $\sim6700$ km s$^{-1}$
H$\alpha$ emission appears largely unchanged from 1993, we find a significant
narrowing of the double-peaked emission profiles in the [\ion{O}{1}]
$\lambda\lambda$6300, 6364 and [\ion{O}{2}] $\lambda\lambda$7319, 7330 lines.  A
comparison of late-time optical spectra of a few other Type II supernovae which,
like SN~1979C, exhibit bright late-time X-ray, optical, and radio emissions,
suggests that blueshifted double-peaked oxygen emission profiles may be a common
phenomenon. Finally, detection of a faint, broad emission bump centered around
5800 \AA\ suggests the presence of WC type Wolf-Rayet stars in the supernova's
host star cluster.

\end{abstract}


\section{INTRODUCTION}

SN 1979C was one of the brightest Type II supernovae (SNe) ever observed in the
optical and radio.  It was discovered in M100 on 1979 April 19 by G.\ Johnson
\citep{Mattei79} (NGC 4321; d = $15.2$ Mpc, \citealt{Freedman01}), a galaxy
which has had an unusually high number (five) of supernovae \citep{Panagia80}.
SN 1979C's blue absolute magnitude was $\approx -20$ mag, about $2-3$ mag
brighter than typical supernovae of its kind in the optical, and it reached a 6
cm flux density of $\sim 8$ mJy, implying a luminosity more than 200 times
that of Cas~A in the radio \citep{Y89,Tammann90,Gaskell92,Weiler86,Weiler89}.

Optical spectra near maximum light showed a featureless continuum that
developed emission lines with expansion velocities  $\ga 8000$ \kms\ typical of
Type II SNe  \citep{Branch81,Barbon82}.  Dominating the spectrum after one
month was strong and broad  (v~$\approx 9000-10,800$ \kms) H$\alpha$ emission
with little or no P Cyg absorption. The relatively slow decline and evolution
of the light curve classified the supernova photometrically as a Type II-L
\citep{Panagia80}.  

SN 1979C was optically recovered in 1990 over a decade after outburst
(\citealt{FM93}; hereafter FM93). Low-dispersion optical spectra showed broad
but blueshifted, double-peaked and asymmetric emission lines of [\ion{O}{1}]
$\lambda\lambda$6300, 6364 and [\ion{O}{2}] $\lambda\lambda$7319, 7330. Faint
6000 \kms\ broad H$\alpha$ emission with an asymmetric profile stronger toward
the blue was also seen.  Follow-up ground-based optical spectra taken in 1993
with the Multiple Mirror Telescope (MMT) and near-UV spectra ($2200-4500$ \AA)
obtained in 1997 with the Faint Object Spectrograph (FOS) aboard the {\sl
Hubble Space Telescope} ({\sl HST}) showed these same features as well as
blueshifted, double-peaked emission lines of C~II] $\lambda\lambda$2324, 2325,
[\ion{O}{2}] $\lambda$2470, \ion{Mg}{2} $\lambda\lambda$2796, 2803 and
[\ion{O}{3}] $\lambda$4363 (\citealt{Fesen99}; hereafter F99).  The blueward
peak was much stronger than the less blueshifted peak in the near-UV lines,
which was interpreted as due to extinction within the expanding ejecta.      

Radio flux density from SN 1979C has been extensively monitored over the last
three decades, with observations beginning eight days after maximum optical
light.  These data show an initial decline in flux density followed by a
flattening or possible brightening at an age of $\approx10$ yr
\citep{Weiler91,Weiler01,Montes00}. The spectral index at late times has been
relatively constant, although recent data suggest a probable flattening
\citep{Montes00,Bartel08}.  Potential quasi-periodic variations in its radio
emission may be due to modulations of the progenitor star's circumstellar medium
(CSM) by a binary companion \citep{Weiler92,Schwartz96}.  Very Long Baseline
Interferometry (VLBI) observations suggest a near-free expansion and shell
structure \citep{Bartel03,Bartel08}.

Here we present a recent optical spectrum of SN 1979C and compare it against
earlier spectra in order to follow the evolution of its late-time emission.
Similarities between SN~1979C and a number of other Type II supernovae observed
at late times ($> 5$ yr) are also briefly discussed.

\begin{figure*}
\centering 
\includegraphics[width=0.9\linewidth]{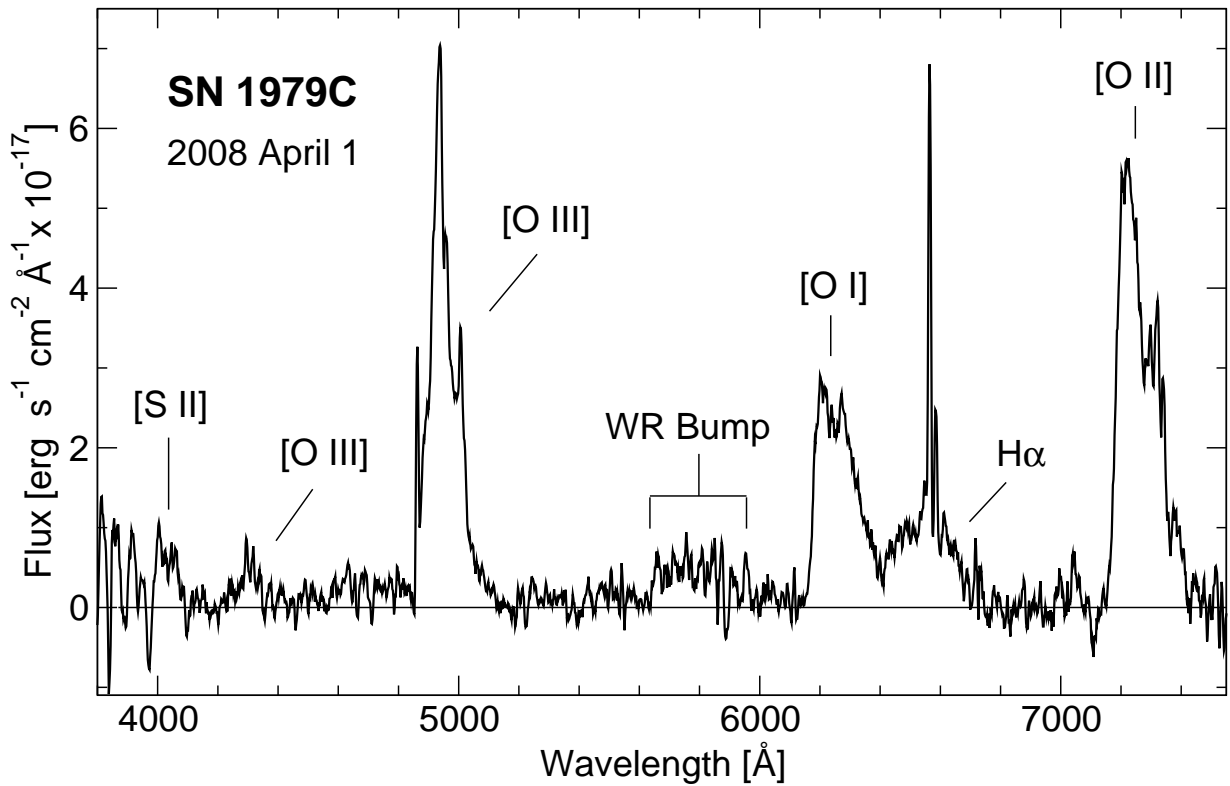}
\caption{MMT spectrum of SN 1979C taken in April 2008.  A blue continuum has
been removed from the observed spectrum. }
\end{figure*}

\section{OBSERVATIONS}

Encouraged by the success of exploratory spectra taken at MDM Observatory in
December 2007, low-dispersion optical spectra of SN 1979C were obtained with
the 6.5~m MMT on Mount Hopkins, Arizona, using the Blue Channel spectrograph on
2008 April 1.  A 1$''$ wide slit and a 300 lines mm$^{-1}$ 4800 \AA\ blaze
grating was used to obtain spectra spanning $3500-7500$ \AA\ with a resolution
of $\sim7$ \AA.  A total of $4 \times 1200$~s exposures were taken at the
parallactic angle under good but non-photometric seeing ($\approx0.7''$).
Spectra were reduced and calibrated employing standard techniques in
IRAF\footnote{IRAF is distributed by the National Optical Astronomy
Observatories, which are operated by the Association of Universities for
Research in Astronomy, Inc., under cooperative agreement with the National
Science Foundation.}. A strong blue continuum likely due to the star cluster in
the vicinity of SN 1979C was removed in the final reduction.

\section{RESULTS}

Our 2008 spectrum of SN 1979C taken some 29 years after maximum light is shown
in Figure 1 and plotted in the rest frame of M100 ($V = 1571$ \kms;
\citealt{Rand95}).  Broad H$\alpha$ emission along with broad and multi-peaked
lines of forbidden oxygen are visible in the spectrum.  Narrow, unresolved
lines are emission from an \ion{H}{2} region local to the SN 1979C site. Below
we discuss the main emission features observed. All listed wavelengths have
been corrected to the rest frame of M100. 

H$\alpha$ --- Faint, asymmetric H$\alpha$ emission spanning approximately $6410
- 6710$ \AA\ ($-7000$ to $+6700$ \kms) is seen centered around the narrow H~II
nebular line at 6563 \AA.  Because the broad H$\alpha$ emission merges toward
the blue with [\ion{O}{1}] $\lambda\lambda$6300, 6364 emission and toward the
red with the [\ion{S}{2}] $\lambda\lambda$6716, 6731 lines, the true maximum
expansion velocities are uncertain. The MMT spectrum was
taken on a mainly clear but not photometric night.  Consequently, we have used a
December 2007 MDM spectrum to estimate line fluxes.  Our estimated flux for the
broad H$\alpha$ emission is $\sim2 \times 10^{-15}$ ergs cm$^{-2}$ s$^{-1}$, a
$\sim$ 35\% decrease from the $(3 \pm 0.5) \times 10^{-15}$ ergs cm$^{-2}$
s$^{-1}$ reported by F99.

\begin{figure*}[htp]
\begin{minipage}{0.475\linewidth}
\centering
\includegraphics[width=\linewidth]{f2.eps}
\caption{Optical spectra of SN 1979C spanning ages $11.1-29.0$ yr. Approximate
  ages are with respect to an estimated optical maximum on 1979 April 15
  \citep{Panagia80}.} 
\end{minipage}
\hspace{0.63cm}
\begin{minipage}{0.5\linewidth}
\centering
\vspace{0.2cm}
\includegraphics[width=\linewidth]{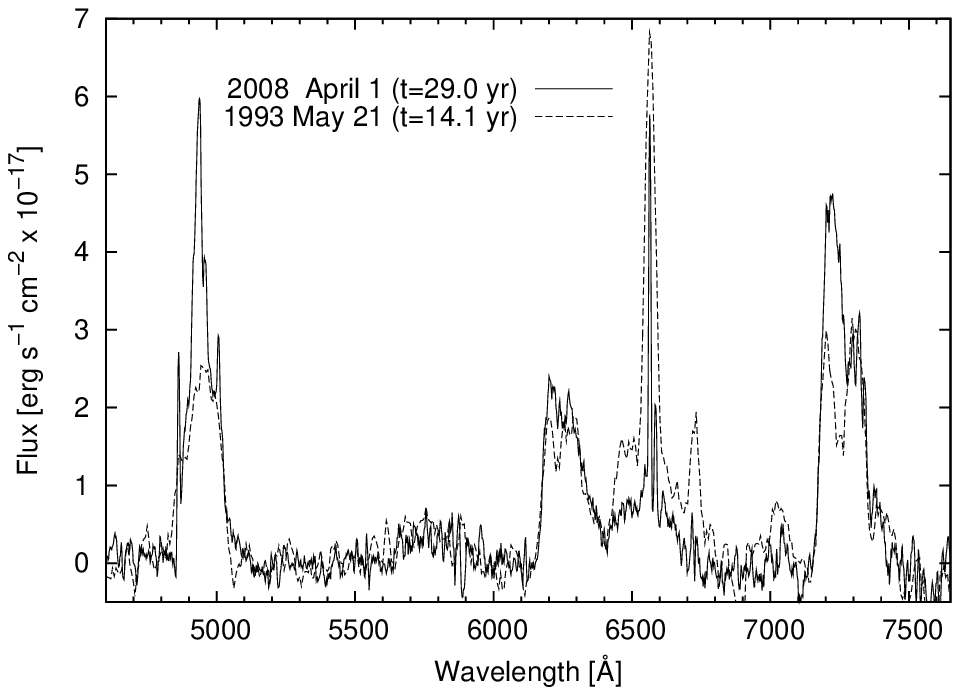}
\caption{MMT optical spectra of SN 1979C at epochs t = \\ 14.1~yr and 29.0 yr.}
\end{minipage}
\end{figure*}

[\ion{O}{1}] --- Like that seen in 1991 and 1993 (FM93, F99), the [\ion{O}{1}]
$\lambda\lambda$6300, 6364 line profile is strongly double-peaked (see Fig.\
1).  We measure peaks at $6210$ and $6275$~\AA, corresponding to expansion
velocities of $-4300$ and $-1200$ \kms\ with respect to 6300~\AA.  The
separation between the peaks in the 2008 spectrum is less than that seen in
1993 when they were at $-4900$ and $-600$ \kms\ (F99; see discussion in \S
4.1). On the other hand, the full expansion velocity of the [\ion{O}{1}] line
profile is $-7000$ to $+1600$ \kms, some 1000 \kms \ larger than that reported
by F99. This difference might simply be attributable to the higher S/N and
greater sensitivity of the more recent MMT observation.  

[\ion{O}{2}] --- The [\ion{O}{2}] $\lambda\lambda$7319, 7330 lines also show a
broad, blueshifted double-peaked emission profile spanning a wavelength range
of $7160-7420$ \AA.  With respect to 7325 \AA\ (the approximate weighted center
of the two emission lines), this corresponds to an expansion velocity range of
$-6700$ \kms \ to $+3900$ \kms.  The profile shows two blueshifted emission
peaks with one at 7220 \AA \ ($-4300$ \kms) and a weaker peak near 7305 \AA \
($-200$ \kms).  Similar double peaks were reported in F99 but at somewhat
different velocities. Also, our new spectrum shows the less blueshifted peak is
now weaker in strength compared to the more blueshifted peak.  Although some
emission from other lines such as [\ion{Ca}{2}] $\lambda\lambda$7291, 7324 is
possible, the  similarities of the double-peaked profile across all oxygen
lines suggests this emission is primarily from [\ion{O}{2}].

[\ion{O}{3}] --- The strong emission observed around 4960~\AA\ is blueshifted
[\ion{O}{3}] $\lambda\lambda$4959, 5007 line emission.  The improved S/N of the
2008 MMT spectrum compared to earlier observations shows the [\ion{O}{3}]
profile clearly for the first time.  With respect to the [\ion{O}{3}]
$\lambda$5007 line, a strong emission peak at 4935 \AA \ corresponds to a
velocity of $-4300$ \kms, which matches the blueshifted emission peak seen in
the other oxygen lines.  Measured to the red of 5007 \AA, faint emission
extends out to 5125 \AA \ corresponding to a velocity of $+7000$ \kms.  This
velocity is significantly larger than $+1600$ \kms\ measured in F99, but, as
noted earlier, the difference is likely attributable to the higher S/N and
greater sensitivity of the 2008 observation.  The narrow, unresolved line at
5007 \AA\ is [\ion{O}{3}] emission from the local M100 \ion{H}{2} region, as is
the narrow H$\beta$ line at 4862 \AA.

We identify faint emission near 4300 \AA \ as blueshifted ($-4300$ km s$^{-1}$)
[\ion{O}{3}] $\lambda$4363 line emission. Assuming $E$($B-V$) = 0.34 (F99), the
observed dereddened I(4959+5007)/I(4363) ratio is $\approx 17$, implying
electron densities of around $10^{5}$ cm$^{-3}$ assuming a temperature of
$25,000$ K.  This ratio is much larger than the value of $\simeq 4$ reported by
F99, suggesting an appreciable decrease in density in the SN's O-rich ejecta
over the last 15 years.

{\it Broad emission at 5800 \AA.} --- Broad emission centered around 5800 \AA\
and extending from $5670-5900$ \AA\ can be seen in Figure 1.  Though faint, this
emission is consistent with earlier reports of broad emission at
$5700-5900$~\AA\ (FM93, F99).  The position, broadness, and strength resembles
the Wolf-Rayet (WR) emission ``bump'' due to \ion{C}{3} $\lambda$5696,
\ion{C}{4} $\lambda\lambda$5801, 5812 and \ion{He}{1} $\lambda$5876 observed in
galaxies and clusters of stars having a population of evolved WR stars mainly of
the WC type \citep{Sidoli06}.

{\it Other emission lines}.---Emission around 4050 \AA\ is identified with
blueshifted [\ion{S}{2}] $\lambda\lambda$4069, 4076, which is consistent with
{\sl HST} FOS spectra obtained in 1997 (F99). The lack of broad [\ion{S}{2}]
$\lambda\lambda$6716, 6731 emission lines implies that electron densities for
the S-rich ejecta lie above $10^{4}$ cm$^{-3}$.

\section{DISCUSSION}

\subsection{Evolution of Late-Time Optical Emission}

Our spectrum shows that SN 1979C remains remarkably optically bright almost
three decades after outburst.  Its late-time X-ray, optical, and radio
emissions are likely the result of interaction with a dense and extensive CSM
environment left behind by the progenitor star
(\citealt{Chevalier94,Weiler86,Weiler89}, FM93, \citealt{Immler05}).  In this
scenario, the broad late-time H$\alpha$ emission is due to shocked
high-velocity H-rich ejecta, while broad oxygen lines are reverse shock heated
O-rich ejecta. 

Strong interactions between supernovae and CSM environments many years after
outburst have been observed in a number of supernovae. Of these, SN 1979C is
notable in that it is one of the few that we have been able to follow its
optical evolution for almost three decades after outburst.  

In Figure 2, we show a comparison of optical spectra of SN 1979C spanning ages
$11-29$ yr after outburst. Overall, these spectra do not show dramatic changes
in the overall strength of the emission lines.  This is
similar to that seen in SN~1979C's X-ray flux which shows no significant
decline since SN 1979C was first detected in 1995 \citep{Immler05}.

However, noticeable changes in the relative strength of the emission lines are
observed. These changes are highlighted in Figure 3, which is an overplot of the
two MMT observations of SN 1979C at t=14 and t=29 yr. Although the flux for each
observation is only accurate within $\pm$20\%, a decline in the ratio of
H$\alpha$/[\ion{O}{1}] emission is apparent, which agrees with our estimated
$\sim$ 35\% decline in H$\alpha$ emission from 1993. The plot also illustrates
how the [\ion{O}{2}] and [\ion{O}{3}] line profiles display an increase in
blueshifted emission centered about $-4300$ \kms\ such that the blueshifted
emission dominates over emission near zero velocity at the present epoch.

\begin{figure*}
\centering
\includegraphics[width=0.8\linewidth]{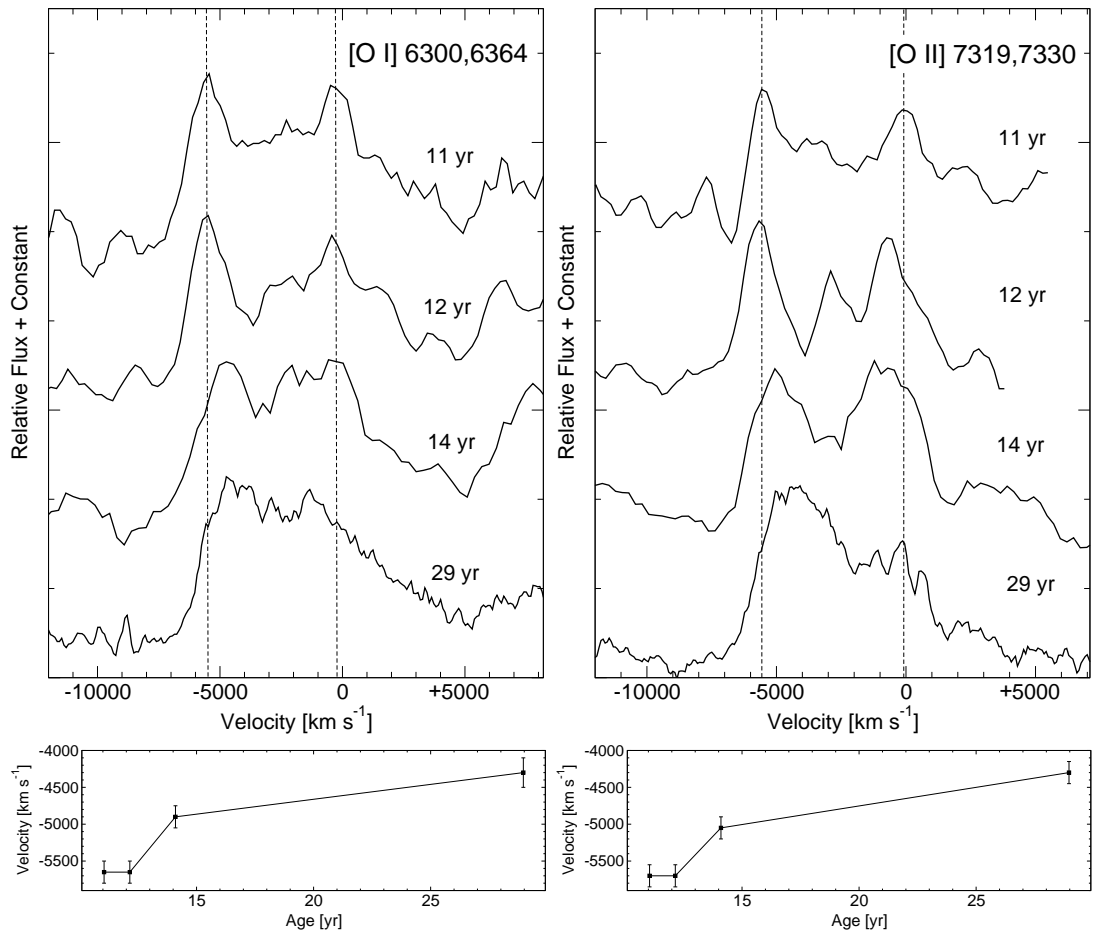}
\caption{Evolution of forbidden oxygen emission line profiles of SN
  1979C. {\sl Top}: The [O I] $\lambda\lambda$6300, 6364 and [\ion{O}{2}]
  $\lambda\lambda$7319, 7330 lines at four epochs. Velocities are with
  respect to 6300 and 7325 \AA\ in the rest frame of NGC 4321. The dashed
  vertical lines mark the velocities of the emission peaks at the earliest
  epoch. {\sl Bottom}: Plots of the velocity of the emission peak of greatest
  blueshift against approximate age of SN~1979C.} 
\end{figure*}

The position of the blueshifted emission peaks in the forbidden oxygen lines
also appears to have shifted over time.  In Figure 4 we show the emission line
profiles of the [\ion{O}{1}] $\lambda\lambda$6300, 6364 and [\ion{O}{2}]
$\lambda\lambda$7319, 7330 lines at all four epochs. As seen in this figure,
although the total width of the line profiles has not changed significantly,
the separation between the two blueshifted emission peaks appears to have
decreased with time.  That is, the more blueshifted emission peak in both the
[\ion{O}{1}] and [\ion{O}{2}] line profiles has moved toward less blueshifted
velocities between 1991 to 2008 (i.e., ages 11 to 29 yr).  On the other hand,
while a wavelength change for the lower velocity blueshifted peak around
$-1000$ \kms \ is more uncertain, the [\ion{O}{1}] line profile peaks in the
2008 spectrum suggest an increased blueshift.

Included along the bottom of Figure 4 are the estimated velocities of the more
blueshifted [\ion{O}{1}] and [\ion{O}{2}] peaks ({\it left and right panels})
plotted against the age of the supernova. These plots suggest a sharp drop in
the velocity of the bluer emission peak between 1990 and 1993 and then a
relatively slow decline from 1993 to 2008. During the last 18 years, the blue
peak has shifted to the red by some $1400$ \kms.  

A similar shift over time of the bluer [\ion{O}{2}] emission peak to lower
velocities was also observed in SN~1986J \citep{Milisavljevic08}. In that case,
the phenomenon was attributed to the progression of the reverse shock inward
toward slower moving O-rich ejecta. The same situation may apply to SN 1979C.
Interpreted in this way, the decrease in velocity of the bluer emission peak
from 5700 km s$^{-1}$ at $t = 11.1$ yr to 4300 km s$^{-1}$ at $t = 29.0$ yr
represents the advance of the reverse shock and can be used to estimate the
ejecta density profile. If the density of the supernova ejecta, $\rho _{\rm
ej}$, as a function of radius $r$ follows the form $\rho _{\rm ej} \propto
r^{-n}$, the velocity of the ejecta at the reverse shock, $v$, should depend on
time $t$ as $v \propto t^{-1/(n-2)}$ provided $n>5$ (\citealt{Chevalier03} and
references therein).  The $1400$ \kms\ shift over epochs $11.1-29.0$ yr years
implies a value of $n = 5.5$, consistent with this scenario but only marginally
so.  A greater value of $n=6.8$ is implied if only the 1993 and 2008 MMT data
are considered.

\subsection{Blueshifted Oxygen Emission}

Blueshifted oxygen emission profiles with one or more emission peaks like those
observed in SN~1979C appear to be a common and long-lasting phenomenon in the
spectra of several Type II SNe which exhibit bright late-time X-ray, optical,
and radio emissions.  Table 1 lists three other Type II SNe which have
late-time oxygen emission lines displaying two blueshifted emission peaks like
those seen in SN 1979C.  As can be seen in the table, these four SNe show
blueshifted emission peaks, one around a velocity of $-200$ to $-1000$ \kms,
and a second, bluer peak at $-3300$ to $-5700$ \kms. Finding four Type II SNe
with strongly blueshifted, double-peaked oxygen emissions is interesting and
may point to a common, late-time emission property. 

On the other hand, some other Type II SNe show quite different oxygen emission
profiles from those of SN~1979C, indicating a range of late-time emission
properties of O-rich ejecta.  For example, SN~1993J at $t \sim 5$ yr showed a
`horned' profile in its [\ion{O}{1}], [\ion{O}{2}], and [\ion{O}{3}] lines with
strong redshifted emission peaks roughly matching the velocity of the
blueshifted peaks \citep{Matheson00a,Matheson00b}.  Alternatively, other
late-time Type II SNe exhibit three or more oxygen emission peaks at both blue
and redshifted velocities (e.g.\ SN~1996cr; \citealt{Bauer08}) or just
blueshifted oxygen emission lacking high-velocity emission peaks (e.g.\
SN~1957D and SN~1986E; \citealt{Capp95}).   

Bridging this diversity of late-time emission properties is the indication that
strongly blueshifted line profiles appear to be common in late-time, Type II SN
spectra even when double or multiple emission peaks are not present. The
commonality of strongly blueshifted oxygen emission in radio and optically
bright Type II SNe suggests that they may share similar properties in the
distribution of their reverse shock heated O-rich ejecta due to interaction
with the surrounding CSM environment, and implicates the presence of dust
blocking emission from the rear expanding hemisphere.

We note that spectra of core-collapse supernovae just entering the nebular
phase ($t > 100^{\rm d}$) often show double-peaked emission profiles in the
lines of [\ion{O}{1}] $\lambda\lambda$6300, 6364 \citep{Modjaz08,Maeda08}.
Such profiles have been interpreted as due to ejecta expanding with a torus or
disk-like geometry. However, these double-peaked profiles sometimes have both
blue and redshifted velocities centered about the [\ion{O}{1}] $\lambda$6300
line.  Thus, the double-peak emission phenomenon in the late-time spectrum of
SN 1979C and other Type II SNe appears different and may not be directly
related. 
\begin{deluxetable}{lcrrll}
\centering
\tablecaption{Type II SNe with Late-time Oxygen Emission Peaks}
\tablecolumns{65}
\tablewidth{0pt}
\tablehead{\colhead{Supernova}                  &
           \colhead{Age}                        &
           \multicolumn{2}{c}{Emission Peaks }  &
           \colhead{Lines observed}             \\
           \colhead{}                           &
           \colhead{(yr)}                       &
           \multicolumn{2}{c}{(\kms)}           &
           \colhead{ [\ion{O}{1}] [\ion{O}{2}] [\ion{O}{3}]} }
\startdata
SN 1970G$^{1}$ & 22 & $-5700$ & \nodata   &  ~~X               \\
SN 1979C$^{2}$ & 12 & $-5700$ & $-200$    &  ~~X ~~~~X ~~~~ X  \\
               & 14 & $-4900$ & $-600$    &  ~~X ~~~~X ~~~~    \\
               & 29 & $-4300$ & $-1000$   &  ~~X ~~~~X ~~~~ X  \\
SN 1980K$^{3}$ & 12 & $-3300$ & $-700$    &  ~~X ~~~~X ~~~~    \\
               & 16 & $-3500$ & $-1000$   &  ~~X ~~~~X ~~~~    \\
SN 1986J$^{4}$ & 6  & $-3500$ & $-1000$   &  ~~X ~~~~X ~~~~ X  \\
               & 24 & \nodata & $-600$    &  ~~X               \\
\enddata
\tablerefs{
        $^{1}$\citet{Fesen93}.
        $^{2}$\citet{FM93,Fesen99}.
        $^{3}$\citet{Fesen94,Fesen99}.
        $^{4}$\citet{Leibundgut91,Milisavljevic08}.
        }
\end{deluxetable}

\subsection{The Progenitor's Stellar Environment}

Lastly, we address the significance of the suspected WR star emission `bump' in
the SN~1979C spectrum.  {\sl HST} imaging shows the presence of a star cluster
in the vicinity of the supernova \citep{vanDyk99,Immler05}.  As can be seen in
Figure 1, our spectrum suggests the presence of a broad emission bump centered
around 5800~\AA\ resembling the so-called `yellow' WR emission feature seen in
WC type stars. The lack of strong He II $\lambda$4686 emission in the spectrum
is also consistent with  WC rather than WN type WR stars in the SN~1979C star
cluster. 
 
Several metal-rich \ion{H}{2} regions in M100 have been observed to possess WR
stars \citep{Pindao02}.  Analysis of the {\sl HST} imaging data shows the
SN~1979C cluster contains young blue stars of ages $4-6$ Myr, consistent with
the ages of WC stars \citep{vanDyk99,Massey03}. Photometry of the cluster's
stellar population yielded an estimate of the progenitor star's initial mass to
be $(17-18) \pm 3$ M$_{\odot}$. Assuming the SN~1979C progenitor was among the
more massive stars in its star cluster, our detection of WR-type emission
features from this cluster is consistent with such high mass estimates for the
SN~1979C progenitor.

\section{CONCLUSIONS}

We present an optical spectrum of SN 1979C taken almost three decades after
outburst. This spectrum is compared against archival spectra to follow the
evolution of SN~1979C's late-time optical emission.  Overall, H$\alpha$
emission appears to have decreased by some 35\% while the [\ion{O}{1}],
[\ion{O}{2}], and [\ion{O}{3}] lines show noticeable increases in blueshifted
emission from levels observed in 1993.  The separation between emission peaks
in the line profiles of [\ion{O}{1}] $\lambda\lambda$6300, 6364 and
[\ion{O}{2}] $\lambda\lambda$7319, 7330 has decreased over the past 18 years,
with the more blueshifted peak's velocity changing from $-5700$ \kms \ to
$-4300$ \kms.  We note that SN~1979C's late-time oxygen line emissions appear
similar to other Type II SNe observed $5 - 30$ yr past outburst in that they
show largely blueshifted emission, often with a blueshifted, double-peaked line
profile. The commonality of these late-time emissions may signal a shared
geometry in the oxygen ejecta and/or CSM environment for X-ray, optical, and
radio bright core-collapse supernovae.

\acknowledgements

We thank C. Gerardy for assistance with the MDM observations, and R. Chevalier
and P. Massey for helpful comments on an earlier draft. R.~P. acknowledges that
supernova research at Harvard is supported by NSF Grant AST06-06772, and
D.~M. thanks NSERC for partial support of this research through a PGS award.


\begin{thebibliography}{}

\bibitem[Barbon et al.(1982)]{Barbon82} Barbon, R., Ciatti, F., Ortolani, S.,
         Rafanelli, P. \ 1982, A\&A, 116, 43
\bibitem[Bartel \& Bientenholz(2008)]{Bartel08} Bartel, N., \&
         Bientenholz, M.~F. \ 2008, \apj, 682, 1065
\bibitem[Bartel \& Bientenholz(2003)]{Bartel03} Bartel, N., \&
         Bientenholz, M.~F. \ 2003, \apj, 591, 301
\bibitem[Bauer et al.(2008)]{Bauer08} Bauer, F.~E., et al. \ 2008, preprint
  (astro-ph/0804.3597)
\bibitem[Branch et al.(1981)]{Branch81} Branch, D. et al. \ 1981,
         \apj, 244, 780
\bibitem[Cappellaro et al.(1995)]{Capp95} Cappellaro, E., 
         Danziger, I.~J., \& Turatto, M.\ 1995, \mnras, 277, 106 
\bibitem[Chevalier \& Fransson(2003)]{Chevalier03} Chevalier, R.~A.,
 \& Fransson, C.\ 2003, Supernovae and Gamma-Ray Bursters, ed K.\ Weiler
         (Berlin: Springer) 598, 171
\bibitem[Chevalier \& Fransson(1994)]{Chevalier94} Chevalier, R.~A., 
          \& Fransson, C.\ 1994, \apj, 420, 268 
\bibitem[Fesen et al.(1999)]{Fesen99} Fesen, R.~A., et al. \ 1999,
         \aj, 117, 725, F99
\bibitem[Fesen \& Matonick(1994)]{Fesen94} Fesen, R.~A., \& Matonick, D.M. \
         1994, \apj, 428, 157 
\bibitem[Fesen \& Matonick(1993)]{FM93} Fesen, R.~A., \& Matonick,
         D.~M. \ 1993, \apj, 407, 110, FM93
\bibitem[Fesen(1993)]{Fesen93} Fesen, R.~A. \ 1993, \apjl, 413, L109
\bibitem[Freedman et al.(2001)]{Freedman01} Freedman, W.~L., et al. \
         \apj, 533, 47
\bibitem[Gaskell(1992)]{Gaskell92} Gaskell, C.~M.\ 1992, \apj, 389, L17
\bibitem[Immler et al.(2005)]{Immler05} Immler, S., et al. \ 2005, \apj, 632, 283
\bibitem[Leibundgut et al.(1991)]{Leibundgut91} Leibundgut, B., et al.\ 1991,
         \apj, 372, 531
\bibitem[Maeda et al.(2008)]{Maeda08} Maeda, K., et al. \ 2008,
         Science, 319, 1220
\bibitem[Massey(2003)]{Massey03} Massey, P. \ 2003, \araa, 41, 15
\bibitem[Matheson et al.(2000a)]{Matheson00a} Matheson, T.~M., et al. \
         2000a, \aj, 120, 1487
\bibitem[Matheson et al.(2000b)]{Matheson00b} Matheson, T.~M.,
         Filippenko, A.~V., Ho, L.~C., Barth, A.~J., \& Leonard, D.~C. \
         2000b, \aj, 120, 1499
\bibitem[Mattei(1979)]{Mattei79} Mattei, J. \ 1979, IAU Circ., No.\ 3348
\bibitem[Milisavljevic et al.(2008)]{Milisavljevic08} Milisavljevic,
         D., Fesen, R.~A., Leibundgut, B., \& Kirshner, R.~P. \ 2008, \apj, in press
\bibitem[Modjaz et al.(2008)]{Modjaz08} Modjaz, M., Kirshner, R.~P.,
         Blondin, S., Challis, P., \& Matheson, T.\ 2008, preprint
	 (astro-ph/0801.0221)
\bibitem[Montes et al.(2000)]{Montes00} Montes, M.~J., Weiler, K.~W.,
         van Dyk, S.~D., Panagia, N., Lacey, Ch.~K., Stramek, R.~A., \& Park,
          R. \ 2000, \apj, 532, 1124
\bibitem[Panagia et al.(1980)]{Panagia80} Panagia, N., et al.\ 1980, \mnras,
         192, 861 
\bibitem[Pindao et al.(2002)]{Pindao02} Pindao, M., Schaerer, D.,
         Gonz\'{a}lez Delgado, R.~M., \& Stasinsky, G. 2002, A\&A, 394, 443
\bibitem[Rand(1995)]{Rand95} Rand, R.~J. \ 1995, \aj, 109, 2444
\bibitem[Schwartz \& Pringle(1996)]{Schwartz96} Schwartz, D.~H., \&
         Pringle, J.~E. \ 1996, \mnras, 282 1018
\bibitem[Sidoli, Smith, \& Crowther(2006)]{Sidoli06} Sidoli, F., Smith, L.~J.,
         \& Crowther, P.~A. \ 2006, \mnras, 370, 799
\bibitem[Tammann \& Schr\"oder(1990)]{Tammann90} Tammann, G.~A., \&
         Schr\"oder, A. \ 1990, A\&A, 236, 149
\bibitem[van Dyk et al.(1999)]{vanDyk99} van Dyk, S.~D., et al. \
         1999, \pasp, 111, 313
\bibitem[Weiler et al.(1992)]{Weiler92} Weiler, K.~W., van Dyk, S.~D.,
         Pringle, J., \& Panagia, N. \ 1992, \apj, 399, 672
\bibitem[Weiler et al.(1991)]{Weiler91} Weiler, K.~W., van Dyk, S.~D.,
         Panagia, N., Sramek, R., \& Discenna, J. \ 1991, \apj, 380, 161
\bibitem[Weiler et al.(1986)]{Weiler86} Weiler, K.~W., Sramek, R.~A.,
         van der Hulst, J.~M., \& Salvatli, M. \ 1986, \apj, 301, 790
\bibitem[Weiler et al.(1989)]{Weiler89} Weiler, K.~W., Panagia, N.,
         Sramek, R.~A., van der Hulst, J.~M., Roberts, M.~S., \& Nguyen, L. \
         1989, \apj, 336, 421
\bibitem[Weiler et al.(2001)]{Weiler01} Weiler, K. W., Panagia, N.,
         Sramek, R.~A., van Dyk, S.~D., Montes, M.~J., \& Lacey, Ch. K. \
         2001, in Supernovae and Gamma-Ray Bursts, eds.\ M.\ Mivio, N.\
         Panagia, \& K.\ Sahu (Cambridge: Canbridge Univ.\ Press), 198
\bibitem[Young \& Branch(1989)]{Y89} Young, T.~R., \& Branch, D. \ 1989, \apj,
         342, L79 

\end{thebibliography}
\end{document}